\documentclass[11pt]{article}
\usepackage{moriond,epsfig}

\def\be{\begin{equation}}
\def\ee{\end{equation}}
\def\bea{\begin{eqnarray}}
\def\eea{\end{eqnarray}}

\begin{document}
\vspace*{4cm}
\title{Q-PYTHIA --- A MONTE CARLO IMPLEMENTATION FOR JET QUENCHING}

\author{ N\'ESTOR ARMESTO$^{(1)}$, LETICIA CUNQUEIRO$^{(2)}$ and CARLOS A. SALGADO$^{(1)}$}

\address{$^{(1)}$ IGFAE and Dep. F\'\i sica de Part\'\i culas, Universidade de Santiago de Compostela (Galicia-Spain)\\
$^{(2)}$ INFN, Laboratori Nazionali di Frascati (Italy)}

\maketitle\abstracts{RHIC experiments have established jet quenching as a fundamental tool in the study of hot matter in heavy-ion collisions. The energy reached is enough for high-$p_T$ inclusive particle studies and one- or two-particle correlations up to $p_T\simeq 10-20$ GeV. The corresponding data are successfully described by a formalism based on radiative energy losses. This formalism needs, however, to be improved for more exclusive observables, in particular for the reconstructed jet properties. Jet reconstruction is a new branch in heavy-ion collisions with very high expectations as a tool for the characterization of the medium properties. We present here an implementation of the jet quenching phenomena based on a modification at the level of the splitting functions in parton showers in standard, general purpose, Monte Carlos as PYTHIA or HERWIG.}

Jet physics is a recent subject in heavy-ion collisions. Indeed, RHIC (and partially also the SPS) have shown in the last ten years how rich the physics studies at the largest possible transverse momentum of the produced particles can be \cite{RHIC}. Most of those data correspond to inclusive one- or two-particle measurements, where trigger bias effects appear. These effects can be generically understood as a geometric bias in which only those particles close to the surface of the produced medium can escape it. Reconstructed jets have long been regarded as ideal tools to surmount this difficulty: in a perfect jet reconstruction no energy is "lost" and the bias associated to triggering in a steeply-falling $p_T$ spectrum disappears. The main difficulty in such studies is to deal with the large multiplicity environment of a heavy-ion collision in which the underlying event represents $\sim 100$ GeV per unit area at RHIC --- and $\sim 250$ GeV is expected at the LHC --- i.e. figures similar to the total energy of the eventually reconstructed jets. This difficult analyses benefitted, however, from tools initially developed to control de pileup in the LHC proton-proton programme \cite{Cacciari:2005hq,Cacciari:2008gn} which found a natural application in heavy-ion collisions. In this way, reconstructed jets at RHIC were first presented at the Hard Probes conference in 2008 \cite{Putschke:2008wn,Salur:2008hs} and, latter on and with important improvements, at Quark Matter 2009 and subsequent conferences. Full control over different experimental biasses and procedures is under ongoing studies by the experimental collaborations. These works  represent the beginning of a completely new avenue in the study of heavy-ion collisions. 

On the theory side, the description of experimental data in terms of the so-called {\it radiative energy loss} is very successful --- see e.g. Ref. \cite{reports} for recent reports on the subject --- although some issues are still open, the most important one being the suppression of heavy quarks in the hot medium. All phenomenological implementations rely, in one way or the other, on the independent gluon emission approximation, first proposed in Ref. \cite{Baier:2001yt} precisely to deal with the bias effects described above. This prescription is enough to describe inclusive one- or two-particle production due to the fact that the medium-induced radiative spectrum is not divergent. However, the way to describe more exclusive observables as jet structures or intra-jet correlations is not known from first principles calculations. Several attempts exist to deal with this problem, in most cases by Monte Carlo implementation \cite{Borghini:2005em,Zapp:2008gi,Renk:2008pp,Lokhtin:2008xi,Armesto:2008qh,qherwig}. 

We present here a modification of the PYTHIA parton shower routine \cite{Sjostrand:2006za} to include the final-state medium-modified parton shower evolution \cite{Armesto:2008qh}. A similar modification in HERWIG \cite{Corcella:2000bw} is under way \cite{qherwig}.

\section{A medium-modified parton shower evolution}

Gluon radiation, independent except for an ordering variable (virtuality, angle, etc...), leads to an excellent description of jet evolution in the vacuum. The ordering is dictated by the presence of divergencies in the splitting probabilities which are resummed in well-controlled approaches. Monte Carlo generators such as PYTHIA or HERWIG encode this physics in parton shower routines. So, gluon multiplication is the main building block of jet structure formation in the vacuum.

Several different effects can affect the jet evolution in the presence of a medium, among others:

\begin{enumerate}

\item Medium-induced gluon radiation --- which modifies the splitting probability producing broadening, softening of the spectrum and enhancement of the jet multiplicity.

\item Non-eikonal corrections to the splitting probability --- collisional energy loss producing a flow of energy (dominantly) from the fast partons to the medium.

\item Modification of the color structure of the jet evolution --- exchanging color with the medium would lead to a different color structure at the end of the parton shower, modifying the hadronization stage.

\item Role of the ordering variable --- the space-time picture of the jet is irrelevant in the vacuum but not in the medium where an interplay between the extension of the medium and that of the jet could be present.

\end {enumerate}

A complete implementation of all these effects does not exist to date. In our implementations \cite{Armesto:2008qh,qherwig}, we have introduced a medium-induced term in the splitting function --- item 1. above --- and a formation time effect as an effective cut-off of the radiation --- item number 4. above. The implementation of extra mechanisms is under way.

\section{The modified splitting probability}

\begin{figure}
\begin{minipage}{0.3\textwidth}
\begin{center}
\includegraphics[width=0.85\textwidth]{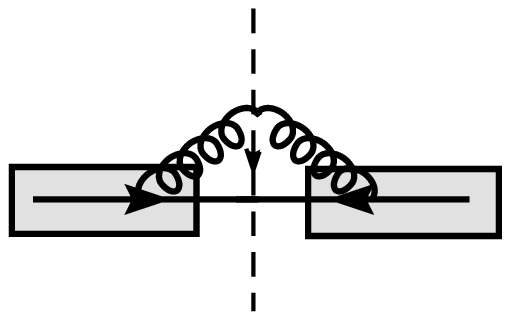}
\end{center}
\end{minipage}
\hfill
\begin{minipage}{0.3\textwidth}
\begin{center}
\includegraphics[width=0.85\textwidth]{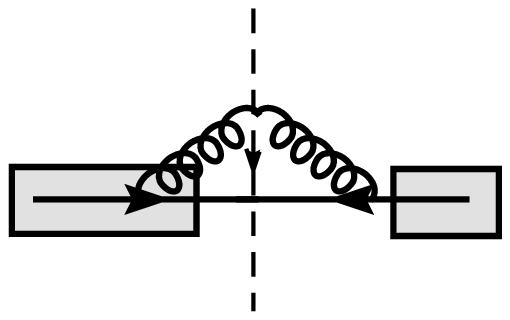}
\end{center}
\end{minipage}
\hfill
\begin{minipage}{0.3\textwidth}
\begin{center}
\includegraphics[width=0.85\textwidth]{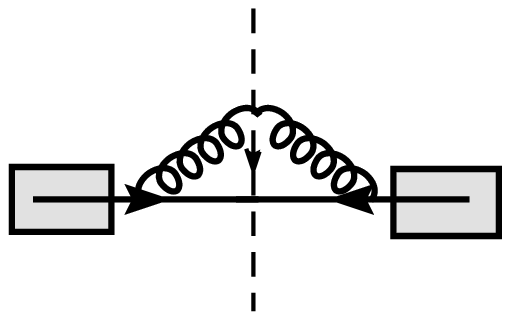}
\end{center}
\end{minipage}
\caption{The three contributions to the gluon radiation in the presence of a medium (represented by a shaded box). The first two are included in the medium-induced gluon radiation term, while the third corresponds to the usual vacuum splitting.}
\label{fig:feynm}
\end{figure}

The central assumption in our approach \cite{msf} is a factorization of the vacuum and medium contributions to the splitting probability $P(z)$ as two independent and additive terms:
 \begin{equation}
P_{\rm tot} (z)= P_{\rm vac} (z)\to
 P_{\rm tot} (z)=P_{\rm vac}(z)+\Delta P(z,t,\hat{q},L,E).
 \label{eq:splitadd}
 \end{equation}
This additivity have been proved at the level of the one-gluon inclusive spectrum, where the total contribution after computing the Feynmann diagrams of Fig. \ref{fig:feynm} can be factorized into a vacuum plus a medium term.
By matching with the vacuum case, the corresponding modification to the splitting probability is written as
\begin{equation}
\Delta P(z,t,\hat{q},L,E)=\frac{2\pi k_\perp^2}{\alpha_s}\frac{dI^{\rm med}}{d\omega dk_\perp^2}
\label{eq:medsplit}
\end{equation}
where  $dI^{\rm med}/d\omega dk_\perp^2$ is given by the medium-induced gluon radiation used previously for RHIC phenomenology \cite{reports}. This spectrum depends on two parameters, the medium length $L$ and the transport coefficient $\hat q$. The latter of these quantities encodes all possible information about the medium properties as temperature, density, etc. in a single parameter, with the interpretation of the average transverse momentum squared that the gluon gets per mean free path in the medium. This simple medium modification is 
 implemented in the standard final-state showering routine PYSHOW in PYTHIA \cite{Sjostrand:2006za}. The corresponding routine is publicly available at the site \cite{qatmc}.

As an attempt to relate virtuality ordering in the PYTHIA parton shower and space-time development, we consider the formation time of the radiated gluons as $t_{\rm form}=2\omega/k_\perp^2$ and subtract this quantity, after each splitting, to the total length of the traversed medium. This mixed approach allows us to implement space-time evolution and treat vacuum and medium radiation on the same footing through the modified splitting probabilities (\ref{eq:splitadd}).

\section{Results}

The first expected effect of including an additive term to the splitting probability is an increase of the jet multiplicity. The typical medium radiation is also broadened in angle and softened in radiated energy with respect to the vacuum case. 
\begin{figure}[h]
\begin{center}
  \includegraphics[width=0.7\textwidth]{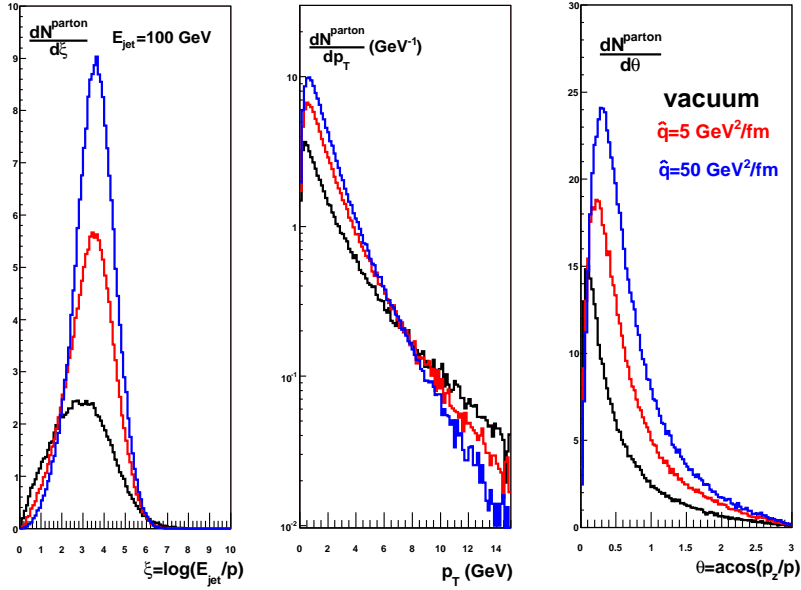}
\end{center}
\caption{Intrajet parton distributions in $\xi$ (left), $p_{T}$ (middle)
 and $\theta$ (right) for a gluon of initial energy $E_{jet}=100$ GeV in a medium of
 length  $L=2$ fm and for different transport coefficients $\hat{q}=0$ (black), 5 (red) and 50 (blue lines) GeV$^2$/fm. Figure from Ref. \protect\cite{Armesto:2008qh}.}
\label{fig:results}
\end{figure}

In Fig. \ref{fig:results} we show the default results at parton level, for two values of the transport coefficient $\hat{q}=5$ and 50 GeV$^2$/fm. We observe a suppression of high-$z$ ($\xi=\log(1/z)$) particles and a big
enhancement of particles with low-intermediate $z$-values, as expected.
We also observe a suppression of high-$p_{T}$
particles and the corresponding enhancement of intermediate-$p_{T}$ particles.
The $p_T$-spectrum should be softer than vacuum at low transverse momentum since low-$p_{T}$ particles should be  kicked towards
higher values of the transverse momentum. However we find a clear enhancement
of low-$p_{T}$ particles. Here, the lack of exchange of energy and momentum
with the medium plus energy conservation in PYTHIA, is making a large effect. 
Finally, we see that the angular distribution broadens with increasing transport
coefficient, as  expected.

\section*{Acknowledgments}

This work has been supported by Ministerio de Educaci\'on
y Ciencia of Spain under projects FPA2005-01963 and FPA2008-01177, by Xunta de Galicia (Conseller\'{\i}a de Educaci\'on), and by the Spanish Consolider-Ingenio 2010 Programme CPAN (CSD2007-00042). 
NA has been supported by  MEC of Spain under a contract Ram\'on y Cajal, and Xunta de Galicia through grant PGIDIT07PXIB206126PR. CAS has been supported by  MEC of Spain under a contract Ram\'on y Cajal, by Xunta de Galicia through grant INCITE08PXIB296116PR and by the European Commission grant PERG02-GA-2007-224770.

\end{document}